\documentclass[aps,twocolumn,prl,showpacs]{revtex4}

\usepackage{amsbsy}
\usepackage{graphicx}

\begin{document}

\newcommand{\beq}{\begin{equation}}
\newcommand{\eeq}{\end{equation}}
\newcommand{\bear}{\begin{eqnarray}}
\newcommand{\eear}{\end{eqnarray}}
\renewcommand{\a}{\alpha}
\renewcommand{\b}{\beta}
\newcommand{\tb}{\tilde{\beta}}
\newcommand{\rd}{\mathrm{d}}
\newcommand{\lp}{\left}
\newcommand{\rp}{\right}
\newcommand{\bs}{\mathbf}
\newcommand{\bz}{\bs{\hat{z}}}
\newcommand{\bx}{\bs{\hat{x}}}
\newcommand{\by}{\bs{\hat{y}}}
\newcommand{\hb}{\hbar}
\newcommand{\g}{\gamma}
\newcommand{\pd}{\partial}
\renewcommand{\l}{\lambda}
\renewcommand{\d}{\delta}
\renewcommand{\o}{\omega}
\renewcommand{\O}{\Omega}
\newcommand{\ofm}{\o_{FM}}
\newcommand{\s}{\sigma}
\renewcommand{\bm}{\bs{M}}
\newcommand{\sm}{\bs{m}}
\newcommand{\bmo}{\bs{M_0}}
\newcommand{\bmp}{\bs{M_\perp}}
\newcommand{\smp}{\bs{m_\perp}}
\newcommand{\bmx}{\bs{M_x}}
\newcommand{\bmy}{\bs{M_y}}
\newcommand{\smx}{\bs{m_x}}
\newcommand{\smy}{\bs{m_y}}
\newcommand{\nb}{\nabla}
\newcommand{\tm}{\times}
\newcommand{\bB}{\bs{B}}
\newcommand{\bb}{\bs{b}}
\newcommand{\bE}{\bs{E}}
\newcommand{\be}{\bs{e}}
\newcommand{\bH}{\bs{H}}
\renewcommand{\sb}{\bs{b}}
\newcommand{\bbz}{\bs{B_z}}
\newcommand{\ba}{\bs{A}}
\newcommand{\bk}{\bs{k}}
\newcommand{\bK}{\bs{K}}
\newcommand{\bQ}{\bs{Q}}
\newcommand{\bh}{\bs{h}}
\newcommand{\sh}{\bs{h}}
\newcommand{\z}{\zeta}
\newcommand{\mxo}{M_{x\,1}}
\newcommand{\mxd}{M_{x\,2}}
\newcommand{\mxt}{M_{x\,3}}
\newcommand{\myo}{M_{y\,1}}
\newcommand{\myd}{M_{y\,2}}
\newcommand{\myt}{M_{y\,3}}
\newcommand{\mxi}{M_{x\,i}}
\newcommand{\myi}{M_{y\,i}}
\newcommand{\mxy}{M_{x,y}}
\newcommand{\oth}{\o_{th}}
\newcommand{\Oth}{\O_{th}}
\newcommand{\lt}{\tilde \l}
\newcommand{\se}{\bs{e}}
\newcommand{\pp}{{(+)}}
\newcommand{\mm}{{(-)}}
\newcommand{\ppm}{{(\pm)}}
\newcommand{\mpm}{m^{(\pm)}}
\newcommand{\bpm}{b^{(\pm)}}
\newcommand{\hpm}{h^{(\pm)}}
\newcommand{\epm}{e^{(\pm)}}
\newcommand{\zpm}{\z^{(\pm)}}
\newcommand{\zr}{\z_{res}^{(+)}}
\newcommand{\im}{\mathrm{Im}}
\newcommand{\re}{\mathrm{Re}}
\newcommand{\D}{\Delta}
\newcommand{\dt}{\mathrm{det}}
\newcommand{\tr}{\textrm}
\setlength\arraycolsep{1pt}

\title{Surface spin waves in superconducting and insulating ferromagnets}
\author{V.~Braude and E.~B.~Sonin}
\affiliation{Racah Institute of Physics, The Hebrew University of
Jerusalem,
Jerusalem 91904, Israel}
\date{\today}
\begin{abstract}
Surface magnetization waves are studied on a semi-infinite magnetic medium
in the perpendicular geometry. Both superconducting and insulating ferromagnets
are considered. Exchange and dipole energies are taken into account,
as well as retardation effects. At large wave vectors, the spectrum for
   a superconductor and insulator  is the same, though for the former the
branch is terminated much earlier than for the latter due to excitation
of plasmons. At small wave vectors, the surface wave is more robust in the superconductor
since it is separated from the bulk continuum by a finite gap.
\end{abstract}
\maketitle 
Studies of spin waves began long ago and proved to be a
powerful method of investigation of magnetically ordered materials
\cite{akhi}. These studies can be divided into two categories. In
the first category, problems are included regarding spin waves in
the bulk of an infinite medium, where long-range geometric effects
do not arise, while the spin-exchange stiffness is taken into
account. The second category deals with spin-wave
modes in finite-size samples of various geometries giving rise to
ferromagnetic resonance (FMR). These modes were usually treated in
the magnetostatic approximation, in which only the long-range
dipole energy is retained in the dynamical equations, while the
exchange energy, as well as well as the retardation effects, are
neglected \cite{damon,akhi}.

  A special class of restricted-geometry spin modes is a surface magnetic
wave, which propagates along the surface of a magnetically ordered
material.   In addition to a great physical interest of surface
modes, they can be important for various technological
applications \cite{miyazaki}, since for these waves the surface serves as a
waveguide, which allows effective spin transport. The latter
attracts now a lot of attention because of the prospects of
spintronics.

The most famous example of a magnetic surface wave is the
Damon-Eshbach wave \cite{damon}, which exists in slabs with the
equilibrium magnetization parallel to the surface (parallel geometry). While  the
existence of this mode can be revealed within the magnetostatic
approximation alone, it is not sufficient to obtain the correct
spectrum. Thus, within this approximation, the surface mode lies
above the bulk modes.  It has been demonstrated \cite{wolfram} 
that inclusion
of the exchange energy can modify the results dramatically. Moreover, in the perpendicular geometry, a surface mode can exist only if the exchange energy is included.
However, the analysis in Ref.~\cite{wolfram} was only
carried out numerically  for some chosen parameter values.
Ref.~\cite{maradudin} considered the effect of displacement
currents (for the parallel geometry), with the conclusion that they also can be 
important, but the exchange energy was disregarded in that work.

This state of affairs, when it is clear that the magnetostatic
approximation is insufficient for the description of the surface
wave spectrum, while no inclusion of the neglected parts has been
done systematically, is part of the motivation for this work.
Another goal of this work is to study the surface-wave spectrum
for a superconducting ferromagnet (SCFM) and to compare its
behavior to that of an insulating ferromagnet and a nonmagnetic
superconductor.
   SCFM's have attracted a lot of attention during the last decade in the
context of unconventional superconductivity. The existence of spin
modes  can serve as a clear proof of magnetic order in these
materials, where detection of ferromagnetism is hindered by
screening Meissner currents \cite{bs}. Knowledge of the
surface-wave properties, while interesting by itself, might be
also useful for a design of new experimental techniques to study
coexistence of ferromagnetism and superconductivity.

This Letter studies the spectrum of surface magnetization waves on
a semi-infinite medium in the perpendicular geometry, taking into
account the dipole and spin-exchange energies and the retardation
effects. Two types of materials are considered: SCFM and
  an insulating ferromagnet. In the SCFM, (longitudinal) plasmons are 
excited together with the transverse magnetic modes
in the surface wave; however, their presence is only important for
 determination of the upper termination point of the branch. On
the other hand, in the insulator, plasmons are
absent, and the surface wave survives to much larger wave vectors. The
retardation effects are important both for very small and very
large wave vectors, the latter case being relevant only for the
insulator, since in the SCFM the branch does not survive to this
regime.
The  spectrum is calculated for both small and large
wave vectors. For large wave vectors, the spectrum is the same for
both types of ferromagnets, apart from the difference in the branch termination
points.
At small wave vectors, the spectra are different for
the two types of materials. The surface branch for the SCFM survives down
to zero wave vector and is separated from the bulk-mode continuum
by a finite gap. The branch for
the insulator is terminated at a very small wave vector, where it
collides with the bulk-continuum bottom.

\begin{figure}[tb]
   \includegraphics[width=0.3\textwidth]{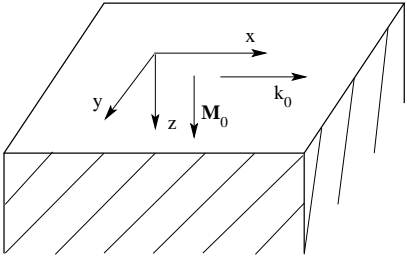}
\caption{Propagation geometry}
\label{setup}
\end{figure}

We consider a semi-infinite uniformly magnetized medium with the
spontaneous magnetization perpendicular to the surface. Physically
this corresponds to a slab whose thickness is large enough that
the other surface can be neglected. Our goal will be to
investigate surface waves in two cases: insulating medium vs. a
superconducting one. We assume no dissipation in both cases, since
only under this assumption, as well as in the semi-infinite
geometry, are surface waves well-defined.
    The setup of
the system is shown in Fig.~\ref{setup}.
The $\bz$ axis together with the spontaneous magnetization $\bmo$ are
   perpendicular to the surface, while the $\bx$ and $\by$ axes  parallel
to it. We also assume a magnetic anisotropy in the system of the
easy-axis type which is strong enough in order to stabilize the
system against the flip of the  magnetization into the $x$-$y$
plane. We will be looking for surface waves propagating along the
$\bx$ direction with a wave vector $k_0$. That is, we will be
interested in solutions of the dynamical equations (specified
below), made of plane-wave
   combinations $\sim \exp(-i\o t+i\bk_i \bs{r})$
   such that $\bk_i=k_0 \bx+q_i \bz$, and $k_0$ is real, while $q_i$ are
complex.

  The magnetization is governed by the Landau-Lifshitz equation, which
gives for a frequency $\o$ \cite{bs}:
\beq \label{eq:motion1}
    -i\o \sm=-g \bmo \tm \sm(\a+\g^2 k^2)+g \bmo\tm \bb,
\eeq
where $\sm$ is the dynamical magnetization such that $\bm=\bmo+\sm$ ($\sm$ is
   in the $x$-$y$ plane); $g$ is the gyromagnetic ratio; $\a$ the magnetic
anisotropy constant (assuming $\a>4\pi$);
   $\g$ the exchange stiffness constant, and, finally, $\sb$ is the magnetic
induction excited by the oscillating magnetization $\sm$. This induction
is related to the magnetization by the generalized London equation, which
is given for plane-wave solutions  by
\beq \label{eq:london}
      \bb=\frac{4\pi k^2 \smp}{k^2+\l^{-2}-K^2}.
\eeq Here  $\l$ is the superconducting penetration depth (for an
insulator, $\l\to \infty$); $K\equiv \o/c$ the electromagnetic
wave vector. The last term $\sim K^2$ in the denominator takes
into account the displacement currents. The vector $\smp$ is the
magnetization component  transverse to the wave vector $\bk$:
\beq
      k^2 \smp=k^2 \sm-\bk (\bk \cdot \sm)=q^2 \sm_x+k^2 \sm_y-q k_0 m_x \bz.
\eeq
Note that when the denominator in the RHS of
Eq.~(\ref{eq:london}) vanishes, this equation does not specify
anymore $\bb$, but, rather, fixes the direction of $\sm$, so that
$k^2\smp=0$. This situation happens for the Damon-Eshbach
wave \cite{damon}. However, in our geometry this possibility can not be
realized, since it would lead to a non-zero $m_z$ component.
Substituting the last two equations into Eq.~(\ref{eq:motion1}), we
obtain a closed equation of motion for the magnetization: 
\[
    i \O \sm = \bz \tm \lp[\lp(\d\a+\g^2k^2+\frac{4\pi\lt^{-2}}{k^2+\lt^{-2}}
    \rp)\sm+\frac{4\pi k_0^2 \sm_x}{k^2+\lt^{-2}} \rp]  ,
\] 
where $\O\equiv \o/gM_0$ and $\lt^{-2}\equiv \l^{-2}-K^2$.
The last term in the RHS breaks the rotational symmetry of the
problem and mixes different circular polarizations. From this
equation, the spectrum of different modes composing the surface
wave is given by
\bear \label{eq:disp}
    \O^2&=&\lp( \d\a+\g^2k^2+\frac{4\pi \lt^{-2}}{k^2+\lt^{-2}}\rp) \nonumber \\
&& \tm\lp(\d\a+\g^2
     k^2+4\pi\frac{k_0^2+\lt^{-2}}{k^2+\lt^{-2}}\rp).
\eear
This is a complicated self-consistent equation in which the
frequency enters both the LHS and the RHS through the definition
of $\lt$.
   For given $\O$ and $k_0$, it gives four
  magnetic modes.

In addition to these magnetic modes, there is also a 
plasma mode in the
system, consisting of a longitudinal electric field alone. Its spectrum is
given by $\o^2=\o_p^2+c_p^2 k^2$
where $\o_p\equiv c/ \l$ and $c_p<<c$ is the plasmon velocity 
originating from the electron gas
compressibility. We will see that
in the SCFM, plasmons are excited together with the magnetic modes 
in the surface
wave.
The electromagnetic (EM) field inside the sample is given by a combination
of the above-mentioned five modes (or four for the insulator). The exact
combination is determined by coupling to the EM field outside the sample
using the appropriate boundary conditions.

The EM field outside the slab can be found using the Maxwell
equations for the vacuum with the frequency $\o$ and wave vector
$\bK=k_0 \bx+k_z \bz$, where $k_z$ is negative imaginary and, as 
before, $K=\o/c$.
The obtained fields should be
coupled to the fields inside the slab by requiring continuity of the EM fields
parallel to the slab surface.
For the insulator,  continuity of
the normal  fields is then satisfied automatically, while for the SCFM,
it is an independent condition which determines the amplitude of the plasmon mode.
Eliminating the latter  from the equations
(for the SCFM), we
obtain from these conditions the following relations for the magnetic
   modes:
\bear \label{eq:cont}
    \sum_i^4 \frac{m_y}{k_i^2+\lt^{-2}}\lp[1+\frac{k_0^2}{q_p k_z}+K^2 \l^2\lp(
     \frac{q}{k_z}-1\rp)\rp]=0 \nonumber \\
    \sum_i^4 \frac{m_x}{k_i^2+\lt^{-2}}\lp[q+\frac{k_0^2+\lt^{-2}}{k_z}\rp]=0,
   \eear
where $q_p=\sqrt{-\lt^{-2}c^2/c_s^2-k_0^2}$
   is the $z$ component of the plasmon wave vector (for the insulator, the term
with $q_p$ should be dropped as the plasmons are irrelevant). In
addition to these, two more conditions are needed for the
amplitudes of spin wave modes. A possible choice for these can be
the condition of vanishing spin currents at the slab surface
\cite{akhi}, which gives
\beq \label{eq:spincur}
\sum_i^4 q_i m_{x,i}=0 \quad \tr{and} \quad
\sum_i^4 q_i m_{y,i}=0\, . \eeq
 These four conditions together with the
requirement that $q_i$ are positive imaginary and $k_z$ negative
imaginary specify a full system of equations whose solution yields
the spectrum of the surface  wave.

We have solved these equations in the limit of small and large
wave vector: $k_0<<\g^{-1}$ and $k_0>>\g^{-1}$ and found
important differences between the SCFM and the insulator in these two regimes.
For a small wave vector, $k_0<<\g^{-1}$, the plasmon contribution in
the SCFM can be neglected
as it is small by the parameter $(c_p/c)^{-1/2}$.
Moreover, in this limit, the spin-wave modes have a definite circular
polarization, and their
dispersion to leading order is given by the usual dispersion of spin
waves in SCFM with $\bk \parallel
   \bz$ \cite{ng, bs}: $  \O=\pm[\d\a +\g^2 q^2+4\pi/(1+q^2 \l^2)]$.
As long as the condition $k_0>>K$ holds
(which is true except for very small wave vectors $\sim 1/C$, where
$C\equiv c/g M_0$ is the light velocity
in magnetic units), retardation effects can be neglected.
   Then 
Eqs.~(\ref{eq:cont}-\ref{eq:spincur}) lead to
\[
    \lp| \begin{array}{cc}
    \frac{1/q_1+\l^2 k_z}{q_1^2+\l^{-2}}-\frac{1/q_2+\l^2 k_z}
    {q_2^2+\l^{-2}} \quad & \frac{1/q_3+\l^2 k_z}{q_3^2+\l^{-2}}-\frac{1/q_4+
     \l^2 k_z}{q_4^2+\l^{-2}} \\
     \frac{1/q_1}{q_1^2+\l^{-2}}-\frac{1/q_2}
    {q_2^2+\l^{-2}} \quad &
     \frac{1/q_3}{q_3^2+\l^{-2}}-\frac{1/q_4}{q_4^2+\l^{-2}}
\end{array}
\rp|=0\,.
\]
In the leading order only the positively polarized modes $(1,2)$
are excited, and the above equation becomes
\beq
    q_1^2+q_2^2+q_1 q_2+\l^{-2}=-\l^2 k_z q_1 q_2 (q_1+q_2)/2\, ,
\eeq
from which a linear surface-wave dispersion is found:
$\O=\O_h+s |k_0|$, where
\beq
    s=\l \frac{(\a-\O_h)\sqrt{1+\g^{-1} \l \sqrt{\a-\O_h}}}{1+
    2\g^{-1} \l\sqrt{\a-\O_h}}
\eeq and $\O_h=\d\a-\g^2 \l^{-2}/2+\g \l^{-1}(4\pi+\g^2
\l^{-2}/4)^{1/2}$. 
This frequency is the frequency of (nonuniform) FMR for a thick SCFM slab \cite{bs2}.
It is remarkable that the surface-wave spectrum contains information about it.

\begin{figure}[tb]
   \includegraphics[width=0.45\textwidth]{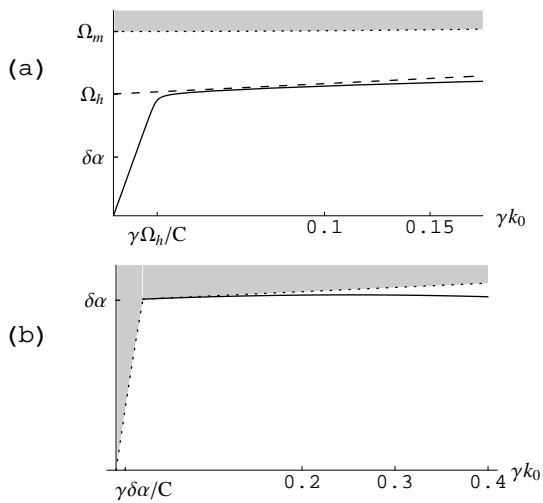}
\caption{Surface wave dispersion $\O(k_0)$ (solid lines) at small $k_0$ (a) for SCFM; (b) for
insulator. The shaded
regions correspond to the bulk-mode continuum. The dashed line in (a)
represents the linear
approximation of the surface-wave spectrum, when the retardation
effects are neglected.}
\label{small}
\end{figure}

At very small wave vectors, $k_0\sim K\sim 1/C$, retardation is
important. When it is taken
into account, the spectrum gets modified, and  instead of the
linear expression, it is given by $\O=\O_h+s(p_z-K^2/p_z)$ where
$p_z^2\equiv k_0^2-K^2$. Solving this, we obtain the SCFM
surface-wave spectrum for
$k_0<<\g^{-1}$:
\beq
    k_0^2=\frac{\D \O^2}{2 s^2}+2K^2+\frac{\D \O}{s}\sqrt{\frac{\D
\O^2}{4s^2}+K^2}.
\eeq
where $\D\O=\O-\O_h$. Its form is given in Fig.~\ref{small}(a). 

As 
$k_0\to0$, only the
long-wavelength mode contributes to the surface wave, and the 
surface-mode spectrum
becomes
$\O=C k_0(1+ \zeta_0^2/2) $, where $\zeta_0$ is the 
surface impedance of
SCFM in the zero-frequency limit \cite{bs}: $\z_0=-i K \l \sqrt{\mu}$ and 
\beq
\mu= \frac{\a(\d\a +\g^2 \l^{-2}+2\g \l^{-1} \sqrt{\a})}{(\d\a+\g \l^{-1}\sqrt{\a})^2}.
\eeq
This is just the surface impedance of a usual superconductor with  magnetic permeability $\mu$,
so the  
magnetic order plays no role in this regime except for adding some effective permeability.
 The corresponding mode, existing on the surface of a conductor, is known as the 
Zenneck wave \cite{Z}. If the medium is not a superconductor, but a metal,
then  the wave is damped.
Note that the bottom of the conduction
band, given at $k_0 \to 0$ by $\O_m=\d\a- \g^2
\l^{-2}+2\sqrt{4\pi}\g \l^{-1}$, is separated from the
surface-wave branch by a finite gap.

For the insulator at $k_0<<\g^{-1}$, the surface-wave frequency  is
$\O\approx \d\a$.
Around this frequency the 
 modes have
the following wave vectors: $k_{1,2}^2\sim
\g^{-1} k_0$;\, $k_3\sim \g^{-1}$, and $k_4\sim K\sim 1/C$. To obtain
the surface wave,
it is enough to keep the first two modes and use them in the two 
equations for $m_y$ from
Eqs.~(\ref{eq:cont},\ref{eq:spincur}).
   The surface-wave spectrum is found from the condition  $q_1=-q_2$,
which  is also the condition
for the bottom of bulk mode continuum. Thus to leading order in
$\g k_0$, these two coincide and are
given by
\beq
    \O=\d\a+\sqrt{8\pi \g^2(k_0^2-2[\d\a/C]^2)}.
\eeq
The limit $C \to \infty $ and $k_0 \to 0$ (while $k_0>>1/C$) produces
$\O=\d\a$, which corresponds to FMR frequency for the insulator.

Taking into account the next order, it is found that the surface wave
is situated slightly
below the
bottom of bulk mode continuum, with a
small separation
$\O_{sw}-\O_{bulk}=\g^2(K^2-k_0^2)\approx-\g^2(k_0^2-\d\a^2/C^2)$.
For very small wave vectors, $k_0 \sim K$,  the retardation effects 
become important.
At this region the branch is terminated, not surviving to $k_0=0$.
 The reason for this is that the bulk-wave continuum at very small $k_0<K$ 
[Fig.~\ref{small}(b)] is
different from that for the superconductor [Fig.~\ref{small}(a)], 
since photons
with velocity $c/\sqrt{\mu_i}$ can propagate in an insulating 
ferromagnet, where $\mu_i=\a/\d\a$. Namely, when $k_0^2=K^2(1+\a/\d\a)$     the
mode $4$ starts propagating. At this point, given by $k_0=k_{0 \,min}^{ins}=
(\a+\d\a)^{1/2}\d\a^{1/2}/C$, the continuum bottom has a cusp, going
down steeply
for smaller $k_0$, where it is no more given by $q_1=-q_2$, but by
$q_4=0$. The
surface-wave branch
collides with the continuum bottom at $k_{0 \, min}^{ins}$  and is
terminated there. Note that near this point the penetration depth of the wave
diverges, so the other surface might become important.
The behavior of the branch  for the insulator at small $k_0$ is shown
in Fig.~\ref{small}(b).

Next we consider the behavior of the surface wave branch for large
$k_0>>\g^{-1}$.
To leading order in $\g k_0$,
   the spectrum can be found using the spin-wave approximation, in 
which only the
exchange stiffness energy is retained, so that the frequency is given
by $\O=\g^2 k^2$ (the  spin-wave
mode
obtained in this approximation will be called mode $1$).
Then the surface-wave spectrum for both the insulator and SCFM is
   found from the condition $q_1=0$, leading to
\beq
    \O=\g^2 k_0^2,
\eeq
which is identical to the continuum bottom. In order to
find a small separation between the
continuum bottom and the surface wave, the  full spin-wave dispersion
relation, Eq.~(\ref{eq:disp})
must be used. Then the wave vector $\bk_1$ has a small imaginary
component perpendicular to
the surface, $q_1=i\pi\g^{-2} k_0^{-1}$, so that
$\O_{sw}-\O_{bulk}=\g^2 q_1^2=\pi^2 \g^{-2}
k_0^{-2}$. These expressions are correct as long as $k_0>>K$, which
is satisfied for the SCFM, as
discussed below. For the insulator, this condition loses its validity
near the (upper) termination
point, and there $q_1=i \pi\g^{-2} (k_0^2-2K^2)/(k_0^2-K^2)^{3/2}$ so
that $\O_{sw}-\O_{bulk}=-\pi^2
   \g^{-2}(k_0^2-2K^2)^2 /(k_0^2-K^2)^3$. While the surface-wave
dispersion is the same for the two
considered types of materials at large $k_0$, the branches are
terminated at points which are very
different. Namely, for the SCFM, the branch termination point is
determined by the condition that
plasmons start propagating. This happens at the plasma frequency,
   when $\lt^{-2} \to 0$, that is, $K^2 \to \l^{-2}$. At this
point, given by $k_0=k_{0\, max}^{SC}=\g^{-1} \O_p^{1/2}$, the
continuum bottom collides with
the surface-wave branch
and terminates it. We see that for $k_0<k_{0\, max}^{SC}$ the
condition $K\le \l^{-1}<<k_0$ is
satisfied, as was stated above.
   In the insulator,
on the other hand, the surface wave
branch survives to  much
larger wave vectors. It is terminated when the mode $1$ starts
propagating, that is, when $q_1=0$,
which gives $k_0=k_{0\, max}^{ins}=\sqrt{2}K=2^{-1/2}\g^{-2}C$.

In conclusion, we have considered a surface wave in 
superconducting and insulating
semi-infinite ferromagnets, taking into account the dipole and the
exchange-stiffness energies, as
well as the displacement currents. The spin-exchange stiffness is
crucial for all regimes. The displacement
currents are important  at very
small wave vectors for both types of materials and very large wave vectors,
near the branch termination point, for the insulator.
The presence of plasmons in the SCFM does not modify the spectrum,
but cuts off the branch at
the plasma frequency.
   For the insulator,
the surface-wave spectrum at very small $k_0$ approaches the continuum
band bottom, whereas
 in the SCFM, the branch remains
separated from the
continuum bottom by a finite gap (which depends on $\g$, $\l$ and $M_0$).
Hence the wave in the latter is more robust against smearing by dissipation.
  The magnetic order in the SCFM makes a profound influence on the
surface wave (except at very small wave vectors $k_0<<K$). Hence the surface wave spectrum 
might provide information about magnetic order in superconductors, in particular, in unconventional superconductors with broken time-reversal symmetry.

We acknowledge discussions with M.~Golosovsky.
This work has been supported by the grant of the Israel
Academy of Sciences and Humanities.

   \end{document}